\documentclass[leqno,twoside]{amsart}

\usepackage{latexsym,esint,amssymb}

\theoremstyle{plain}

\newcommand{\R}{\mathbb R} 
\def\endproof{\hspace*{\fill}\mbox{\ \rule{.1in}{.1in}}\medskip }
\newcommand{\vek}{\vec}

\newtheorem{theorem}{Theorem}[section]

\theoremstyle{definition}

\numberwithin{equation}{section}
\numberwithin{figure}{section}

\begin{document}

\title[nonlinear shell theory with varying thickness]
{A nonlinear theory for shells with slowly varying thickness}
\author{Marta Lewicka, Maria Giovanna Mora and Mohammad Reza Pakzad}
\address{Marta Lewicka, University of Minnesota, Department of Mathematics, 
206 Church St. S.E., Minneapolis, MN 55455}
\address{Maria Giovanna Mora, Scuola Internazionale Superiore di Studi Avanzati,
via Beirut 2-4, 34014 Trieste, Italy}
\address{Mohammad Reza Pakzad, University of Pittsburgh, Department of Mathematics, 
139 University Place, Pittsburgh, PA 15260}
\email{lewicka@math.umn.edu, mora@sissa.it, pakzad@pitt.edu}

\begin{abstract}
We study the $\Gamma$-limit of 3d nonlinear elasticity for shells of small, variable thickness,
around an arbitrary smooth 2d surface.
\end{abstract}

\maketitle

\section{Introduction} \label{introduction}
The following question receives large attention in the current literature 
on elasticity \cite{ciarbook}: which the\-ories of thin
objects (rods, plates, shells) are predicted by the 3d nonlinear theory?
For plates, this problem has been extensively studied through $\Gamma$-convergence;
first by LeDret and Raoult \cite{LR1}, leading to a rigorous derivation of the membrane theory,
and later by Friesecke, James and M\"uller \cite{FJMhier}, for the Kirchhoff, von K\'arm\'an 
and linear theories.  In this framework, much less has been done for shells. 
The membrane and the bending theories were obtained in 
\cite{lr} and \cite{FJMM_cr}, respectively.
More recently, the generalized von K\'arm\'an and linear theories have been
rigorously introduced and justified by the authors in \cite{LMP}.
In this paper, we present these last new results, in an extended version
for shells with variable thickness and mid-surface.

\section{The setting of the problem and the related elastic energy functionals} \label{mainresults}
\noindent {\bf The three-dimensional problem.} 
Let $S$ be a 2d compact, connected, oriented surface in $\mathbb{R}^3$, 
of  class $\mathcal{C}^{1,1}$,
whose boundary is the union of finitely many (possibly none) lipschitz curves.  
Let $\vek{n}(x)$ be the unit normal vector, $T_x S$ the tangent space, and 
$\Pi(x) = \nabla \vek{n}(x)$ the shape operator on $S$, at a given $x\in S$.

For given positive lipschitz functions $g_1, g_2: S\longrightarrow (0,\infty)$,
small in $L^\infty$, consider a family of thin shells:
$$ S^h = \{z=x + t\vek{n}(x); ~ x\in S, ~ -hg_1(x)< t < h g_2 (x)\}, \qquad 0<h< 1.$$
The {\em elastic energy} (scaled per unit thickness) of a deformation 
$u^h\in W^{1,2}(S^h,\R^3)$ is given by:
$$E^h(u^h) = \frac{1}{h}\int_{S^h} W(\nabla u^h),$$
where the stored-energy density $W:\mathbb{R}^{3\times 3}\longrightarrow [0,\infty]$ 
is $\mathcal{C}^2$ in a neighborhood of $SO(3)$, and satisfies:
\begin{equation}\label{assump}
\begin{split}
\forall F\in\mathbb{R}^{3\times 3}\quad \forall R\in SO(3)\qquad
& W(R) = 0, \quad W(RF) = W(F),\\
& W(F)\geq C \mathrm{dist}^2(F, SO(3)). 
\end{split}
\end{equation}
\noindent {\bf The two-dimensional problem.} 
We now introduce the {\em von K\'arm\'an functional} on $S$:
\begin{equation}\label{vonKarman}
\begin{split} 
& I(V,B_{tan})= 
\frac{1}{2} 
\int_S (g_1 + g_2) \mathcal{Q}_2\left(x,B_{tan} - \frac{\kappa}{2} (A^2)_{tan} - 
\frac 12 {\rm{sym}} (A \nabla ((g_2-g_1) \vek{n}) )\right) \\ 
& \qquad \qquad \qquad \qquad + \frac{1}{24} \int_S (g_1 + g_2)^3 
\mathcal{Q}_2\left(x,(\nabla(A\vek{n}) - A\Pi)_{tan}\right),
\end{split}
\end{equation} 
defined for $V\in \mathcal{V}$ and $B_{tan} \in \mathcal{B}$.
The space $\mathcal{V}$ consists of  {\em infinitesimal isometries} 
$V\in W^{2,2}(S,\mathbb{R}^3)$, that is these vector fields $V$ for whom there exists 
a matrix field $A\in W^{1,2}(S,\mathbb{R}^{3\times 3})$ so that:
$$ \partial_\tau V(x) = A(x)\tau \quad \mbox{and} \quad  A(x)^T= -A(x) \qquad 
\forall {\rm{a.e.}} \,\, x\in S \quad \forall \tau\in T_x S.$$
The {\em finite strain space} $\mathcal{B}$ (see \cite{LMP}), consists of  the following 
symmetric matrix fields:
$$\mathcal{B} = \Big\{L^2 - \lim_{h\to 0}\mathrm{sym }\nabla w^h; 
~~ w^h\in W^{1,2}(S,\mathbb{R}^3)\Big\}.$$
In (\ref{vonKarman}) $\kappa\geq 0$ is a parameter, and the positive definite quadratic 
forms $\mathcal{Q}_2(x,\cdot)$ are defined as follows:
$$ \mathcal{Q}_2(x, F_{tan}) = \min\{\mathcal{Q}_3(\tilde F); ~~ (\tilde F - F)_{tan} = 0\}, 
\qquad \mathcal{Q}_3(F) = D^2 W(\mbox{Id})(F,F).$$
The quadratic form $\mathcal{Q}_3$ is defined for all $F\in\mathbb{R}^{3\times 3}$, 
while $\mathcal{Q}_2(x,\cdot)$,  for a given $x\in S$ is defined on tangential minors $F_{tan}$ 
of such matrices.   Recall that the tangent 
space to $SO(3)$ at $\mbox{Id}$ is $so(3)$. As a consequence,  
both forms depend only on the symmetric parts of their arguments
and are positive definite on the space of symmetric matrices \cite{FJMgeo}. 

\medskip

\noindent {\bf Relative stretching and bending.} 
The two terms in (\ref{vonKarman}) are strictly tied to the deformations
of the {\em geometric mid-surface} of $S^h$. 
Given $V\in\mathcal{V}$ and a field $w\in W^{1,2}(S,\mathbb{R}^3)$, consider the deformations:
$$\tilde \phi^h = \mbox{id} + 1/2 h(g_2 - g_1) \vek{n}, \qquad 
\phi^h = \tilde\phi^h + hV + h^2 w.$$
Then $\tilde\phi^h(S)$ is the geometric mid-surface.
A straightforward calculation shows that for any $\tau \in T_x S$:
$$ |\partial_\tau \phi^h|^2 - |\partial_\tau \tilde \phi^h|^2 = 2h^2 \tau^T \Big  
(\mbox{sym} \nabla w 
- 1/2 A^2 - 1/2 \mbox{sym} (A \nabla ((g_2-g_1) \vek{n}))\Big )\tau + \mathcal O(h^3).$$ 
Hence (putting $\kappa=1$ and $B_{tan}=\mbox{sym}\nabla w$), 
the expression in the argument of $\mathcal{Q}_2$ in 
the first term of (\ref{vonKarman}) describes {\em stretching}, namely the second order 
in $h$ change of the first fundamental form of $\tilde\phi^h(S)$.

The second term of (\ref{vonKarman}) relates to {\em bending}, that is the first order 
in $h$ change in the second fundamental form of $\tilde\phi^h(S)$.  
To see this, let $\Pi^h$ be the shape operator on $\phi^h(S)$. 
Indeed, for $\tau\in T_xS$ one obtains:
$$(\nabla\phi^h)^{-1} \Pi^h (\nabla \phi^h)\tau 
- (\nabla\tilde\phi^h)^{-1} \Pi (\nabla\tilde\phi^h)\tau
= h\Big(\partial_\tau (A\vek n) - A\Pi\tau\Big) +  \mathcal O(h^2).$$ 

Notice also, that the term $-1/2 \mbox{sym}(A\nabla((g_2-g_1) \vek{n}) )$ which is new 
with respect to analysis in \cite{LMP}, disappears when $g_1 = g_2$ or equivalently 
$S=\tilde\phi^h(S)$. This term expresses the first order in $h$ 
deficit for $V$ from being an infinitesimal isometry on $\tilde\phi^h(S)$.
It is because for any $\eta=\partial_\tau \tilde \phi^h(x)$ where $\tau\in T_xS$ we have:
$$\partial_\eta (V\circ (\tilde\phi^h)^{-1})\cdot \eta
= \partial_\tau V\cdot\partial_\tau \tilde\phi^h 
= -h/2 \tau^T \mbox{sym} (A\nabla((g_2-g_1) \vek{n}) )\tau.$$

\section{The $\Gamma$-convergence of the elastic energy functionals}
\begin{theorem}\label{liminflimsup}
Let $e^h$ be a sequence of positive numbers, for which we assume:
\begin{equation}\label{ehass}
\lim_{h\to 0} e^h/h^4 = \kappa^2 <\infty.
\end{equation} 
{\bf (a)}
For any sequence of deformations $u^h\in W^{1,2}(S^h,\mathbb{R}^3)$ satisfying: 
\begin{equation}\label{basic}
E^h(u^h)\leq C e^h,
\end{equation} 
there exist a sequence $Q^h\in SO(3)$ and $c^h\in\mathbb{R}^3$ 
such that for the normalized rescaled deformations:
$y^h(x+t\vek{n}) = (Q^h)^T u^h(x+ht\vek{n}) - c^h$ defined on the common domain $S^1$,
the following holds.
\begin{enumerate}
\item[(i)] $y^h$ converge in $W^{1,2}(S^{1})$ to the projection $\pi:S^1\longrightarrow S$
along the normal $\vek n$.
\item[(ii)]  The related scaled average displacements:
\begin{equation*}
(V^h[y^h])(x) = \frac{h}{\sqrt{e^h}} \fint_{-g_1(x)}^{g_2(x)}
y^h(x+t\vek{n}) - (x+ ht\vek{n}) ~\mbox{d}t
\end{equation*} 
converge (up to a subsequence) in $W^{1,2}(S)$ to some $V\in \mathcal{V}$.
\item[(iii)] $\frac{1}{h} \mathrm{sym} \nabla V^h[y^h]$ converge
(up to a subsequence) weakly in $L^{2}(S)$ to some $B_{tan}\in\mathcal{B}.$ 
\item[(iv)] $\liminf_{h\to 0} \frac{1}{e^h} E^h(u^h) \geq I(V,B_{tan}).$
\end{enumerate}
{\bf (b)} 
Conversely, for every $V\in \mathcal{V}$ and $B_{tan}\in\mathcal{B}$, there exists a sequence 
$u^h\in W^{1,2}(S^{h},\mathbb{R}^3)$ satisfying (\ref{basic}) and such that (i), (ii), (iii) hold 
(with $Q^h = \mathrm{Id}$ and $c^h = 0$)
and moreover: $\lim_{h\to 0} \frac{1}{e^h} E^h(y^h) = I(V, B_{tan}).$
\end{theorem}

\medskip

\noindent {\bf Proof.} Here we outline the main steps of the proof of Theorem \ref{liminflimsup}.   
We refer the readers to \cite{LMP} for a detailed exposition  
in the case $g_1=g_2 = const$. All convergences below are up to a subsequence.

\noindent {\bf 1.}
A first major ingredient is approximating $u^h$ by $W^{1,2}$ matrix fields $R^h: S\to SO(3)$,
thanks to the scaling invariant nonlinear quantitative rigidity estimate
from \cite{FJMgeo}.
Noting (\ref{basic}) and the uniform equivalence of the functionals 
$\int \mbox{dist}^2(\nabla\cdot,SO(3))$ and $E^h(\cdot)$ on the uniformly lipschitz
domains $S^h$, it follows that:
$$\|\nabla u^h - R^h\pi\|_{L^2(S^h)}\leq Ch^{1/2}\sqrt{e^h}, \qquad
\|\nabla R^h\|_{L^2(S)} + \|R^h - Q^h\|_{L^2(S)} \leq Ch^{-1}\sqrt{e^h}$$
for some $Q^h\in SO(3)$. 
Further, for some skew-symmetric matrix field $A\in W^{1,2}(S,\mathbb{R}^3)$, one has:
\begin{equation*}
\begin{split}
&h/\sqrt{e^h} ((Q^h)^TR^h - \mbox{Id}) \longrightarrow A \quad \mbox{weakly in }
W^{1,2}(S),\\
&h^2/e^h \mbox{sym}((Q^h)^TR^T - \mbox{Id}) \longrightarrow 1/2 A^2\quad \mbox{strongly in }
L^{2}(S).
\end{split}
\end{equation*}
\noindent {\bf 2.}
Define now: 
$$w^h(x+t\vek n) = h/\sqrt{e^h} ((Q^h)^Tu^h(x+ht\vek n) - c^h - (x+ht\vek n)),$$ 
with $c^h$ such that $\int_{S^1}w^h = 0$.  To prove (i) and (ii), one first uses the above convergences 
to obtain that:
$$\nabla_{tan} w^h \longrightarrow A\pi (\mbox{Id}+t\Pi)^{-1} \quad \mbox{ and } \quad
\partial_{\vek n} w^h \longrightarrow 0, \quad \mbox{strongly in } L^{2}(S^1).$$
By the Poincar\'e inequality, $w^h$ must now converge to $V\circ \pi$, strongly in $W^{1,2}(S^1)$.\\
\indent For (iii), one equates various terms in $\frac{1}{h}  \mathrm{sym} \nabla V^h[y^h]$ 
to notice that it converges weakly in $L^{2}(S)$ to:
\begin{equation}\label{Btan}
\fint_{-g_1}^{g_2} (\mbox{sym } G)_{tan} ~\mbox{d}t + \frac{\kappa}{2} (A^2)_{tan} 
+ \frac{1}{2}\mbox{sym} (A\nabla ((g_2-g_1)\vek n)) =: B_{tan},
\end{equation}
where $G$ is the weak $L^2(S^1)$ limit of the matrix fields:
$$G^h(x+t\vek n) = \frac{1}{\sqrt{e^h}} ((R^h)^T\nabla u^h(x+ht\vek n) - \mbox{Id}).$$
The convergence of $G^h$ \cite{LMP}, which is the $\sqrt{e^h}$ order term in the expansion 
of the nonlinear strain $(\nabla u^h)^T\nabla u^h$ at $\mbox{Id}$, plays a major role 
for the expansion of the energy density: $W(\nabla u^h) = W(\mbox{Id} + \sqrt{e^h} G^h)$. 

\smallskip

\noindent {\bf 3.}
Towards proving (iv), the previously established convergences and the definition of 
$\mathcal{Q}_2$ imply:
\begin{equation}\label{inf}
\begin{split}
& \liminf_{h\to 0}\frac{1}{e^h} E^h(u^h)  \geq \frac{1}{2} \int_S \int_{-g_1}^{g_2} 
\mathcal{Q}_2(x, (\mbox{sym } G)_{tan})~\mbox{d}t\mbox{d}x\\
& = \frac{1}{2} \int_S \int_{-(g_1+g_2)/2}^{(g_1+g_2)/2} 
\mathcal{Q}_2\Big(x, (\mbox{sym } G_0)_{tan} + (s+(g_2-g_1)/2)(\nabla_{tan}A)\vek n\Big)
~\mbox{d}s\mbox{d}x,
\end{split}
\end{equation}
where we used the fact that: 
$$(\mbox{sym } G)_{tan}(x+t\vek n) = (\mbox{sym } G_0)_{tan}(x)
+ t(\nabla_{tan}A)\vek n$$ 
is linear in $t$.  Calculating $(\mbox{sym } G_0)_{tan}$ from 
(\ref{Btan}) and collecting various terms in the argument of the quadratic form $\mathcal{Q}_2$
above, we see that the right hand side in (\ref{inf}) equals the von K\'arm\'an
functional $I(V, B_{tan})$.

\smallskip

\noindent {\bf 4.}
If $V\in \mathcal{V}\cap W^{2,\infty}(S, \R^3)$ and $B_{tan} = \mbox{sym} \nabla w$ with 
$w\in W^{2,\infty}(S, \R^3)$, then the recovery sequence is given by the formula  below.
When $V$ and $B_{tan}$ do not have the required regularity, one proceeds by approximation 
and truncation, as in \cite{LMP}.
\begin{equation*}
\begin{split}
y^h(x+t&\vek{n}) = x + h/2 (g_2 -g_1) \vek{n}(x) + {\sqrt{e^h}}/{h} V(x) + \sqrt{e^h} w(x) \\
&  + h(t - 1/2 (g_2 -g_1)) \vek{n}(x) 
+ {(t- 1/2 (g_2 -g_1))}\sqrt{e^h} \Big(\Pi V_{tan} - \nabla (V\vek{n})\Big)(x)\\
& - h (t- 1/2 (g_2 -g_1)) \sqrt{e^h} \vek{n}^T \nabla w + {(t- 1/2 (g_2 -g_1))}h \sqrt{e^h} d^{0}(x) \\
&  + 1/2{(t - 1/2 (g_2 -g_1))^2} h\sqrt{e^h} d^{1}(x).
\end{split}
\end{equation*}
The vector fields $d^{0}, d^{1}\in W^{1,\infty}(S,\mathbb{R}^3)$ are given by
means of the linear map $c(x, F_{tan})$ which returns the unique vector $c$ satisfying:
$$\mathcal{Q}_2(x,F_{tan}) = \min\{\mathcal{Q}_3(F_{tan} + c\otimes \vek{n}(x) 
+ \vek{n}(x)\otimes c); ~~ c\in\mathbb{R}^3\}.$$ Then:
\begin{equation*}
\begin{split}
d^{0}  = & 2c\left(x, B_{tan} - \frac{\kappa}{2}(A^2)_{tan} 
- \frac{1}{2} \mbox{sym} \left( A\nabla ((g_2-g_1) \vek{n} ) \right) \right) \\
& \qquad\qquad + {\kappa}A^2\vek{n}  - \frac{\kappa}{2} (\vek{n}^T A^2 \vek{n})\vek{n} +
\frac{1}{2} \Big (A\nabla ((g_2 - g_1)\vek n) \Big)^T\vek n\\
d^{1} = & 2c\left(x, \mbox{sym }(\nabla(A\vek{n}) - A\Pi)_{tan}\right)
+ \vek{n}^T A\Pi - \vek{n}^T\nabla(A\vek{n}). \hspace{2cm} \endproof 
\end{split}
\end{equation*}

\section{The dead loads and convergence of minimizers}
Consider a sequence of forces $f^h\in L^2(S,\mathbb{R}^3)$ acting on $S$, 
with the following properties:
\begin{equation}\label{fhass}
\int_S (g_1 + g_2)f^h = 0 \quad \mbox{and} \quad \lim_{h\to 0}\frac{1}{h\sqrt{e^h}} f^h = f 
\quad \mbox{weakly in }  L^2(S).
\end{equation} 
Define their extensions 
$f^h (x + t\vek{n}) = \det\left(\mbox{Id} + t\Pi(x)\right)^{-1} f^h(x)$ over $S^h$,
and the maximized action of $f^h$ over all rotations of $S^h$ as:
$$m^h = \max_{Q\in SO(3)} \frac{1}{h} \int_{S^h} f^h(z) \cdot Q z ~\mbox{d}z.$$
The total energy functional on $S^h$ is:
$$ J^h(u^h) = E^h(u^h) + m^h - \frac{1}{h}\int_{S^h} f^h u^h.$$
As in \cite{LMP}, one can prove that:
$$0\geq \inf\left\{\frac{1}{e^h} J^h(y^h); ~~ u^h\in W^{1,2}(S^h, \mathbb{R}^3)
\right\}\geq -C.$$
We further introduce the relaxation function $r: SO(3)\longrightarrow [0,\infty]$,
with its effective domain $\mathcal{M} = \{\bar{Q}\in SO(3);~r(\bar{Q}) < \infty \}$:
\begin{equation}\label{relax}
r(Q) = \inf \Big \{\liminf_{h\to 0} \frac {1}{e^h}
\Big ( m^h - \frac 1h \int_{S^h} f^h \cdot Q^h z~ \mbox{d} z \Big ); 
~ Q^h \in SO(3),~ Q^h \to Q \Big \}. 
\end{equation}
\begin{theorem}\label{thmain}
Assume (\ref{ehass}) and (\ref{fhass}). Let $u^h\in W^{1,2}(S, \mathbb{R}^3)$
be any  minimizing sequence of $\frac{1}{e^h} J^h$, that is:
\begin{equation}\label{minimizers}
\lim_{h\to 0} \left(\frac{1}{e^h}J^h(y^h) - \inf\frac{1}{e^h}J^h \right)=0,
\end{equation}
Then the conclusions of Theorem \ref{liminflimsup} (a) hold, and 
any accumulation point $\bar Q$ of $\{Q^h\}$ belongs to $\mathcal{M}$.
Moreover, any limit $(V, B_{tan}, \bar Q)$ minimizes the following functional,
over $V\in\mathcal{V}$, $B_{tan}\in\mathcal{B}$ and $\bar Q\in \mathcal{M}$: 
$$ J(V, B_{tan},\bar Q) = I(V, B_{tan}) - \int_S (g_1 + g_2) f\cdot \bar{Q} V + r(\bar{Q}).$$
\end{theorem}
The proof follows as in \cite{LMP}.
An equivalent formulation of Theorem \ref{thmain} in terms of $\Gamma$-convergence is possible.

\medskip

\noindent{\bf Example.} When $f^h = h \sqrt{e^h} f$ and $g_1 = g_2$, then:
$$\mathcal{M} = \left\{\bar{Q}\in SO(3); ~~ \int_S f\cdot \bar{Q}x
= \max_{Q\in SO(3)} \int_S f\cdot Qx \right\}$$ 
(in the general case the
inclusion $\subset$ is still true, but not the other one). 
Further, $r\equiv 0$ on $\mathcal{M}$,
so the term $r(\bar Q)$ may be dropped in the definition of $J$.
In the general case, both $r$ and $\mathcal{M}$ depend on the asymptotic behavior
of the maximizers of the linear functions
$SO(3)\ni Q\mapsto \int_{S^h} f^h(z) \cdot Qz~\mbox{d}z$.

\medskip

\noindent{\bf Approximately robust surfaces and higher scalings.} 
Some classes of surfaces, including surfaces of revolution, developable surfaces 
with no flat part, and elliptic surfaces, have the ability to recompense the second order 
stretching (of $\tilde\phi^h(S)$) introduced by the infinitesimal isometry $V$, 
through a suitable second order displacement. 
As an example, surfaces with flat parts do not enjoy this property (for any $g_1$, $g_2$).
For such surfaces \cite{LMP}, described by condition:
$$\{A^2 + {\rm{sym}} (A \nabla ((g_2-g_1) \vek{n})); ~ V\in\mathcal{V}\}\subset \mathcal{B},$$ 
the limit elastic energy $I(V, B_{tan})$  simplifies 
and should be replaced (in Theorems \ref{liminflimsup} and \ref{thmain}, 
regardless of $\kappa\geq 0$) by:
\begin{equation}\label{beta>4} 
\tilde I(V) = \frac{1}{24} 
\int_S (g_1 + g_2)^3 \mathcal{Q}_2\Big(x,\big(\nabla(A\vek{n}) - A\Pi\big)_{tan}\Big).
\end{equation}
When $\kappa=0$ and $g_1=g_2$, the stretching term is negligible for any $S$ and 
hence the von K\'arm\'an theory again reduces to the linear theory, with the elastic energy 
given in (\ref{beta>4}).
For more discussion, see \cite{LMP}.



\section*{Acknowledgments}
M.L. was partially supported by the NSF grant DMS-0707275.
M.G.M. was partially supported 
by the Italian Ministry of University and Research
through the project ``Variational problems with multiple scales'' 2006 and 
by GNAMPA through the project ``Problemi di riduzione di dimensione
per strutture elastiche sottili'' 2008.

\end{document}